\documentclass{PoS}
\usepackage[utf8]{inputenc}
\usepackage{amsmath}

\newcommand{\Herwig}{\textsf{Herwig}}
\newcommand{\VBFNLO}{\textsf{VBFNLO}}
\newcommand{\matchbox}{\textsf{Matchbox}}

\title{Parton-shower Effects in Vector-Boson-Fusion Processes}

\ShortTitle{Parton-shower Effects in Vector-Boson-Fusion Processes}

\author{\speaker{Michael Rauch}\\
        Institute for Theoretical Physics, Karlsruhe Institute of Technology\\
        E-mail: \email{michael.rauch@kit.edu}}

\author{Simon Plätzer\\
        Institute for Particle Physics Phenomenology, Durham University \\
        School of Physics and Astronomy, University of Manchester}

\abstract{%
We investigate the effects of combining next-to-leading order QCD
results with parton-shower effects in $W^+W^-jj$ production via
vector-boson fusion including leptonic decays. Using the
\textsf{Herwig}~7 framework interfaced to \textsf{VBFNLO}~3, we compare
the predictions obtained from the angular-ordered and dipole-based
parton shower algorithms combined with subtractive, MC@NLO-type, and
multiplicative, Powheg-type, matching. A consistent treatment of
renormalisation and factorisation scale variations in the hard process
and the parton shower allows to assign more reliable theory uncertainty
predictions to key distributions like the central rapidity gap.%
}

\FullConference{XXIV International Workshop on Deep-Inelastic Scattering and Related Subjects\\
		11-15 April, 2016\\
		DESY Hamburg, Germany}

\begin{document}

\section{Introduction}
Production of electroweak bosons via vector-boson fusion and
vector-boson scattering, collectively referred to as VBF in the
following, is one of the main process classes to study during the run-2
phase of the LHC. It proceeds via the space-like exchange of electroweak
bosons between two scattering (anti-)quarks.
The characteristic feature of VBF processes are two energetic jets in
the forward regions of the detector, the so-called tagging
jets~\cite{Zeppenfeld:1999yd}. In the central region between the two
tagging jets, the jet activity is reduced, which can be exploited for
example by a mini-jet veto~\cite{Rainwater:1996ud}. This distinguishes
them from two classes of background processes which form an irreducible
background. These are multi-boson production, where one boson decays
into a quark--anti-quark pair, and QCD-induced production in association
with two jets. The application of tight VBF cuts, typically an invariant
mass of the tagging jets above several hundreds of GeV and a large
rapidity separation between them, reduces the contribution of these
background processes and strongly suppresses interference terms with the
VBF process~\cite{Campanario:2013gea}. Also, interference effects
between $t$- and $u$-channel boson exchange become negligible and
justify using the VBF approximation, where these terms are removed.
VBF processes are an ideal tool to study the gauge structure of the SM
due to the appearance of triple and, in particular, quartic gauge
couplings. Diagrams with quartic and triple gauge vertices as well as
with the exchange of a Higgs boson exhibit a strong cancellation among
them, which would be spoiled if any of them receives anomalous
contributions. Hence, VBF is a sensitive probe to test such effects.

Next-to-leading order (NLO) QCD corrections to all VBF processes have
been
calculated~\cite{Figy:2003nv}.
Their effect is rather modest, with corrections typically up to 10\%.
Taking the momentum transfer through the exchanged bosons as a dynamical
scale proves to be an advantageous choice. All other higher-order
corrections have been established for VBF-Higgs production only so far.
The NLO electroweak
corrections~\cite{Ciccolini:2007jr} have
a similar size as the NLO QCD ones. NNLO
QCD~\cite{Bolzoni:2010xr} and NNNLO QCD~\cite{Dreyer:2016oyx} corrections
to the inclusive cross section in the structure-function approach are
below the percent level and at the 1-2 permill level, respectively. NNLO
QCD corrections to differential distributions in the VBF approximation
show much larger effects~\cite{Cacciari:2015jma} up to 10\%, though at
least for some distributions are modelled reasonably well by
parton-shower effects matched to NLO QCD calculations.
Studies of matching the NLO
QCD calculations to parton showers have been performed for some of the
VBF
processes~\cite{Nason:2009ai,Jager:2013mu}
in the POWHEG-BOX
framework~\cite{Nason:2004rx,Frixione:2007vw}, and for
$W^+W^-jj$ production~\cite{Rauch:2016upa} using
\Herwig~7~\cite{Bellm:2015jjp} and
\VBFNLO~3~\cite{Arnold:2008rz}.
To be able to assess the effects, it is necessary to not only compare
the central predictions, but also to study the associated theory
uncertainties. Their size can be estimated from a variation of the
different scales entering the predictions, but also using different
matching schemes and parton-shower algorithms. We will thereby focus on
the perturbative part of the simulation and not consider any
hadronisation or multi-parton-interaction effects.

\section{Calculational Setup}
We perform our study~\cite{Rauch:2016upa} using the \Herwig~7 Monte
Carlo event generator~\cite{Bellm:2015jjp}, which is based on
\textsf{HERWIG++}~\cite{Bahr:2008pv}. Two different matching schemes are
implemented, namely subtractive, MC@NLO-type
matching~\cite{Frixione:2002ik}, and multiplicative, Powheg-type
matching~\cite{Nason:2004rx}. These can be combined with both parton
shower modules available in \Herwig~7, the dipole
shower~\cite{Platzer:2009jq} and the angular-ordered
shower~\cite{Gieseke:2003rz}. 
Its \matchbox\ module, based on Ref.~\cite{Platzer:2011bc}, performs the
simulation of NLO QCD events. 
The necessary squared amplitudes, both for the Born and the
real-emission process as well as the Born-virtual interference term, are
in our study provided by
\VBFNLO~3~\cite{Arnold:2008rz}, which allows us
to get fast and accurate predictions. The communication between the two
programs is done using the BLHA~2 standard~\cite{Alioli:2013nda}. 
We have extended the interface by the possibility to use the dedicated
phase-space generator of \VBFNLO~3, and use this for the results shown
in the following.
%
As example VBF process we take the electroweak production of
$pp \rightarrow W^+ W^- jj \rightarrow e^+ \nu_e \mu^- \bar{\nu}_\mu jj\,$.
The leptonic decays of the $W$ bosons as well as off-shell, 
non-resonant and Higgs exchange contributions are fully included. For
the partons we employ the VBF approximation and neglect interference
terms between quarks of the same flavour in the final state. 
We perform our study for the LHC at a centre-of-mass energy of $13\ {\rm
TeV}$. As we are not interested in top-quark effects, we apply a perfect
bottom-quark veto on the final state. 
At the analysis level we apply typical VBF selection cuts,
\begin{align}
p_{T,j} &> 30 {\rm GeV} \,, & |y_j| &< 4.5 \,, & 
m_{j1,j2} &> 600 {\rm GeV} \,, & |y_{j1}-y_{j2}| &> 3.6 \,, \nonumber\\
p_{T,\ell} &> 20 {\rm GeV} \,, & |y_{\ell}| &< 2.5 \,, & m_{e^+,\mu^-} &> 15 {\rm GeV} \,. 
\label{eq:vbfww_cuts}
\end{align}
Jets are clustered from massless partons via the anti-$k_T$
algorithm~\cite{Cacciari:2008gp} with a cone radius of $R=0.4$.
As PDF set MMHT2014~\cite{Harland-Lang:2014zoa} is used.

\section{Results}
\begin{figure}
~\hfill
\includegraphics[width=0.4\textwidth]{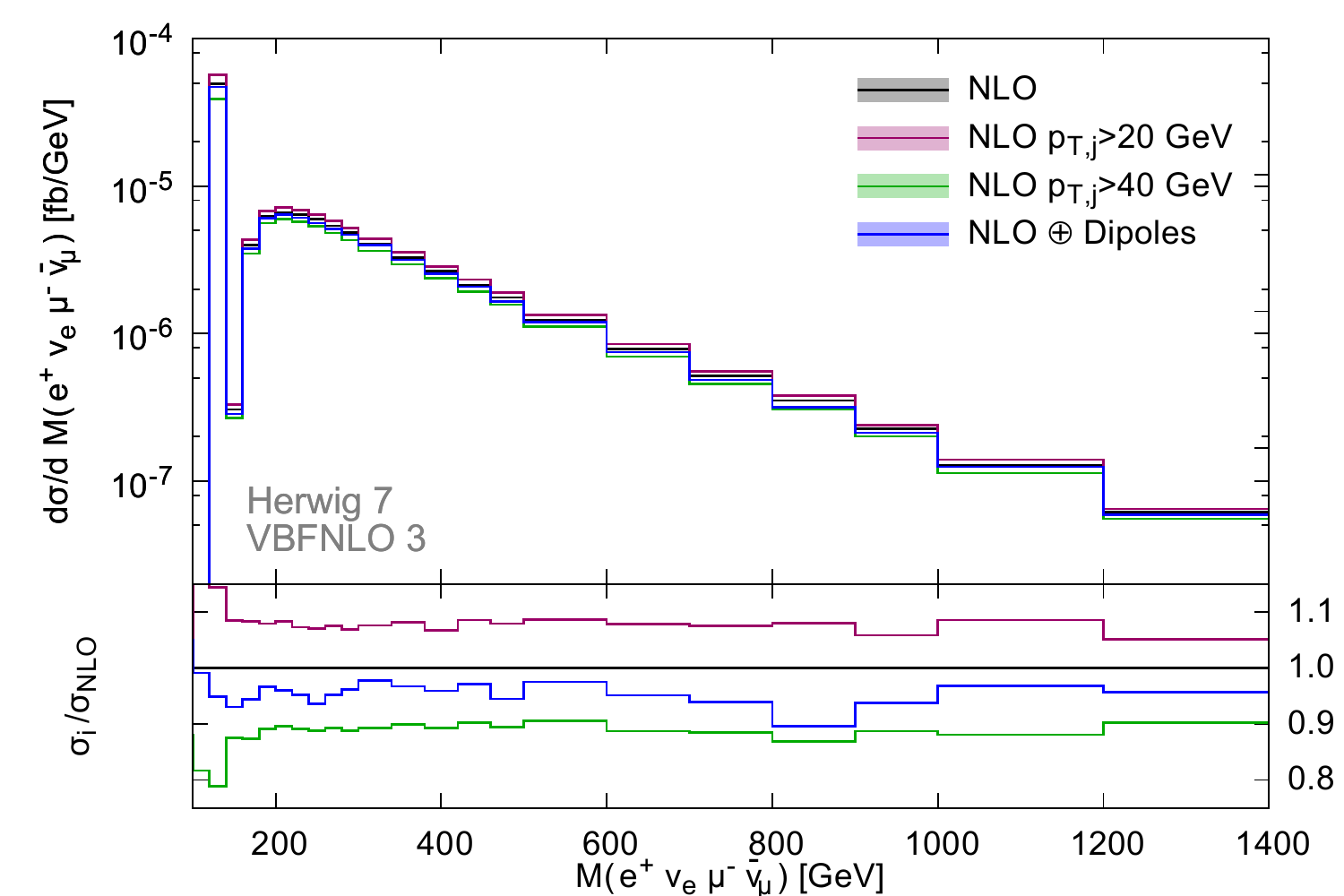}
\hfill\hfill
\includegraphics[width=0.4\textwidth]{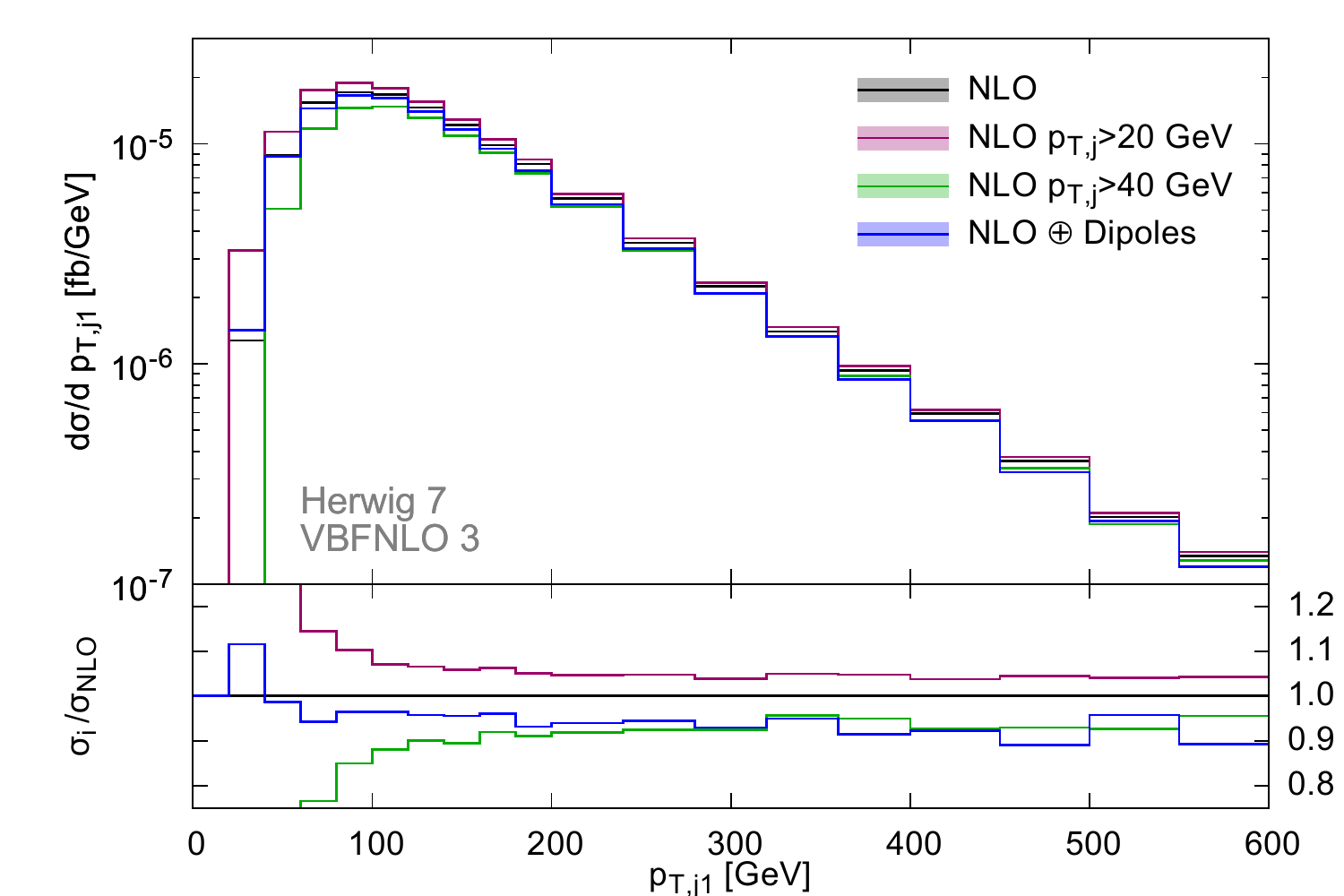}
\hfill~
\caption{Invariant mass of the four-lepton system (\textit{left}) and
transverse momentum of the leading jet (\textit{right}).
Shown is the fixed-order NLO prediction applying different cuts on the jet
transverse momentum and matched with the dipole shower.
Right panel taken from Ref.~\protect\cite{Rauch:2016upa}.
}
\label{fig:migration}
\end{figure}
Parton-shower emissions can change the kinematic properties of an event,
if the emission is sufficiently hard and emitted under a wide angle such
that it is not clustered back into the emitting jet. This in turn can
lead to a migration across cut boundaries, typically reducing the cross
section. How this affects our process is shown in
Fig.~\ref{fig:migration}. Here we show two differential distributions,
namely the four-lepton invariant mass and the transverse momentum of the
leading jet. In both cases, the parton-shower calculation leads to
consistently lower cross sections than the fixed-order result. 
For comparison, we also show fixed-order results where the transverse
momentum cut has been lowered to 20~GeV, the generation-level cut, or
increased to 40~GeV, which is expected to mimic parton-shower effects in
size. We find that the reduction of the cross section is mostly due to
events where additional radiation leads to a reduction of the invariant
mass of the tagging jets, thus more events fail this VBF cut.

\begin{figure}[t]
~\hfill
\includegraphics[width=0.4\textwidth]{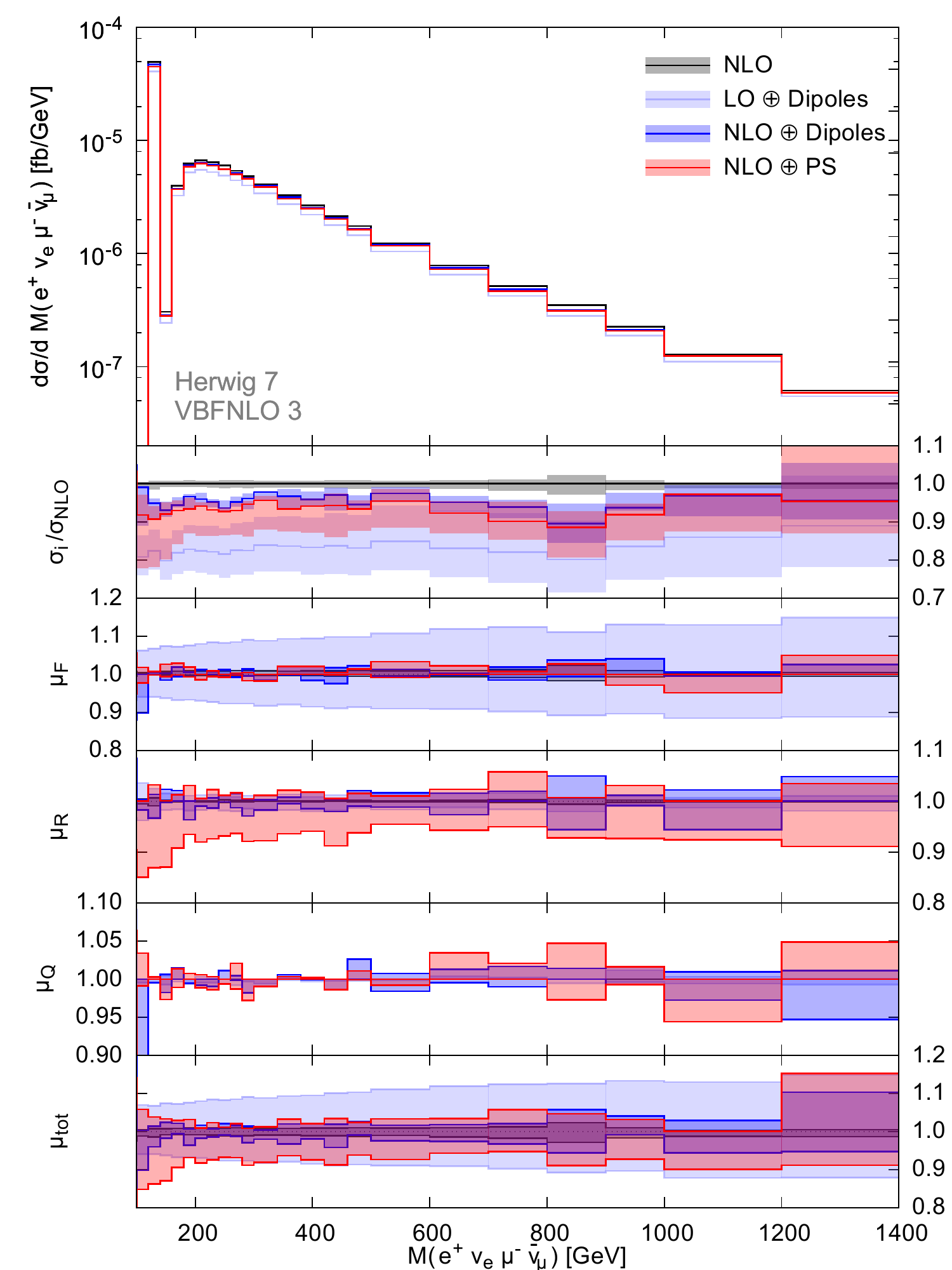}
\hfill\hfill
\includegraphics[width=0.4\textwidth]{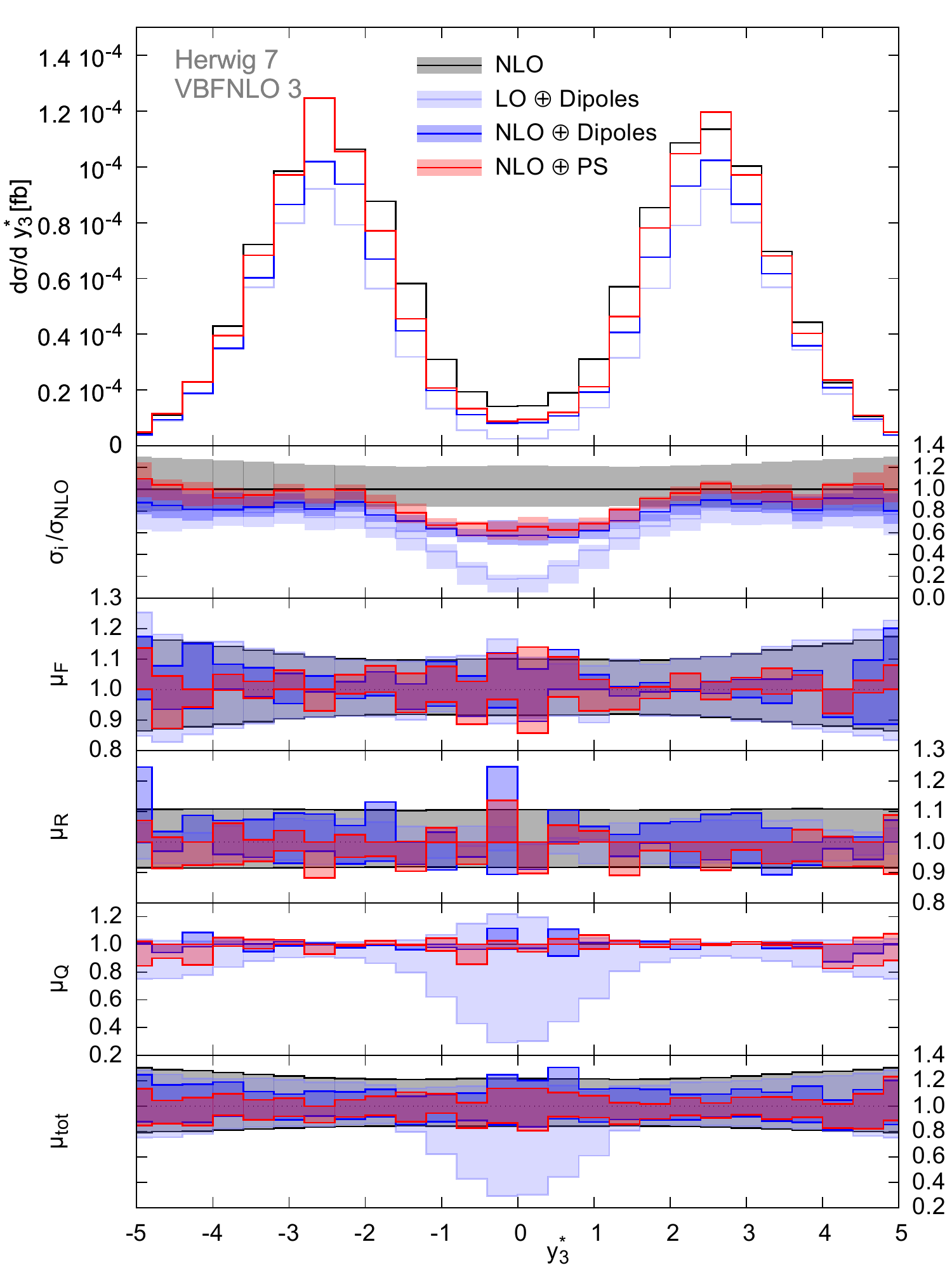} 
\hfill~
\caption{Invariant mass of the four-lepton system (\textit{left}) and
rapidity of the third jet relative to the two tagging jets
(\textit{right}), comparing parton-level NLO results (black), LO plus
dipole shower (light blue), and NLO results combined with the dipole
(dark blue) and angular-ordered (red) shower using the subtractive
matching procedure.
Each panel shows, from top to bottom, the central prediction, central
prediction and overall scale variation normalised to the fixed-order
result, and variation bands for factorisation ($\mu_F$), renormalisation
($\mu_R$), and hard veto ($\mu_Q$) scale as well as the overall envelope
($\mu_{\text{tot}}$) normalised to the respective central prediction.
Figure taken from Ref.~\protect\cite{Rauch:2016upa}.
}
\label{fig:uncert_mcatnlo}
\end{figure}
\begin{figure}[t]
~\hfill
\includegraphics[width=0.30\textwidth]{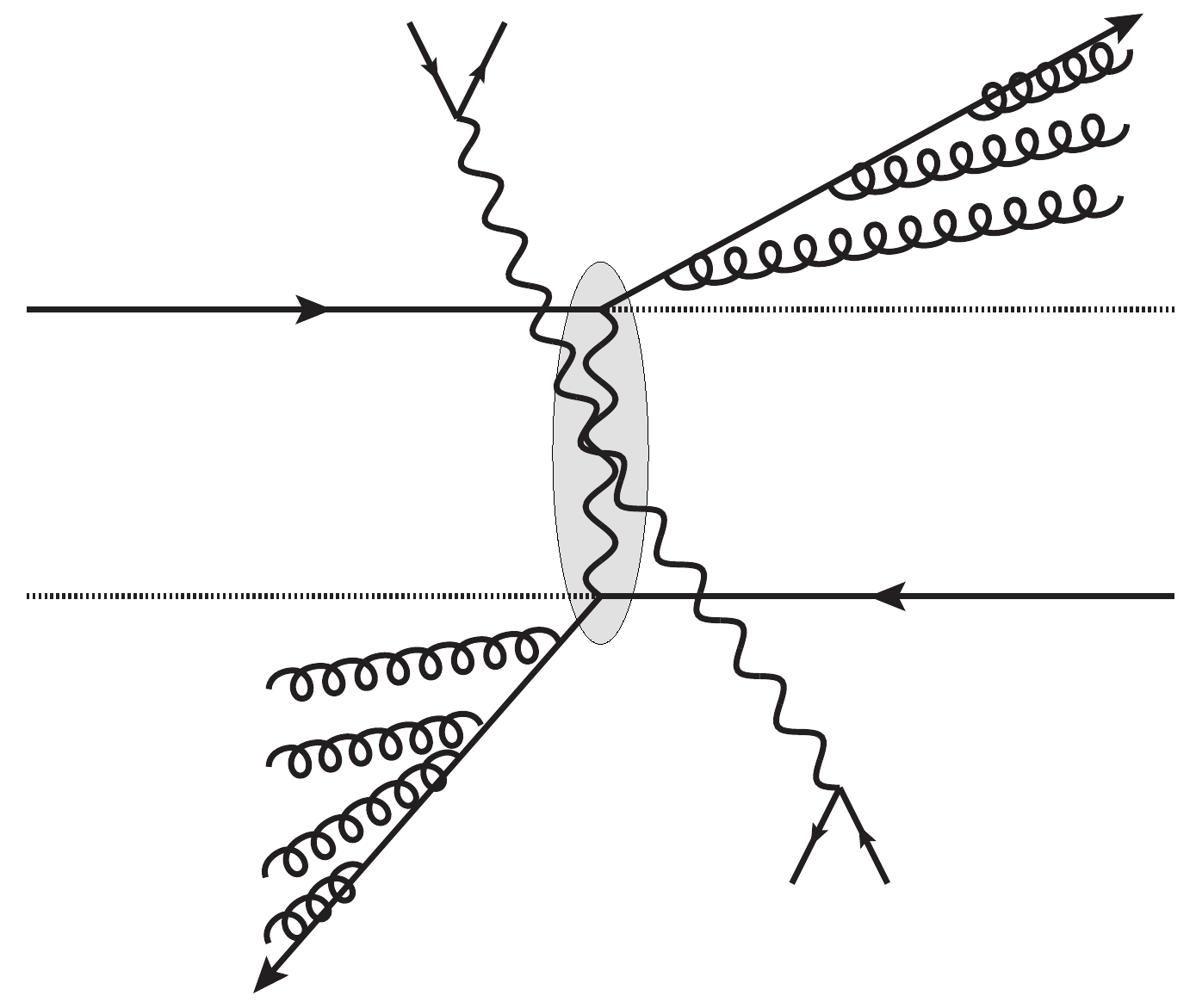} 
\hfill\hfill
\includegraphics[width=0.40\textwidth]{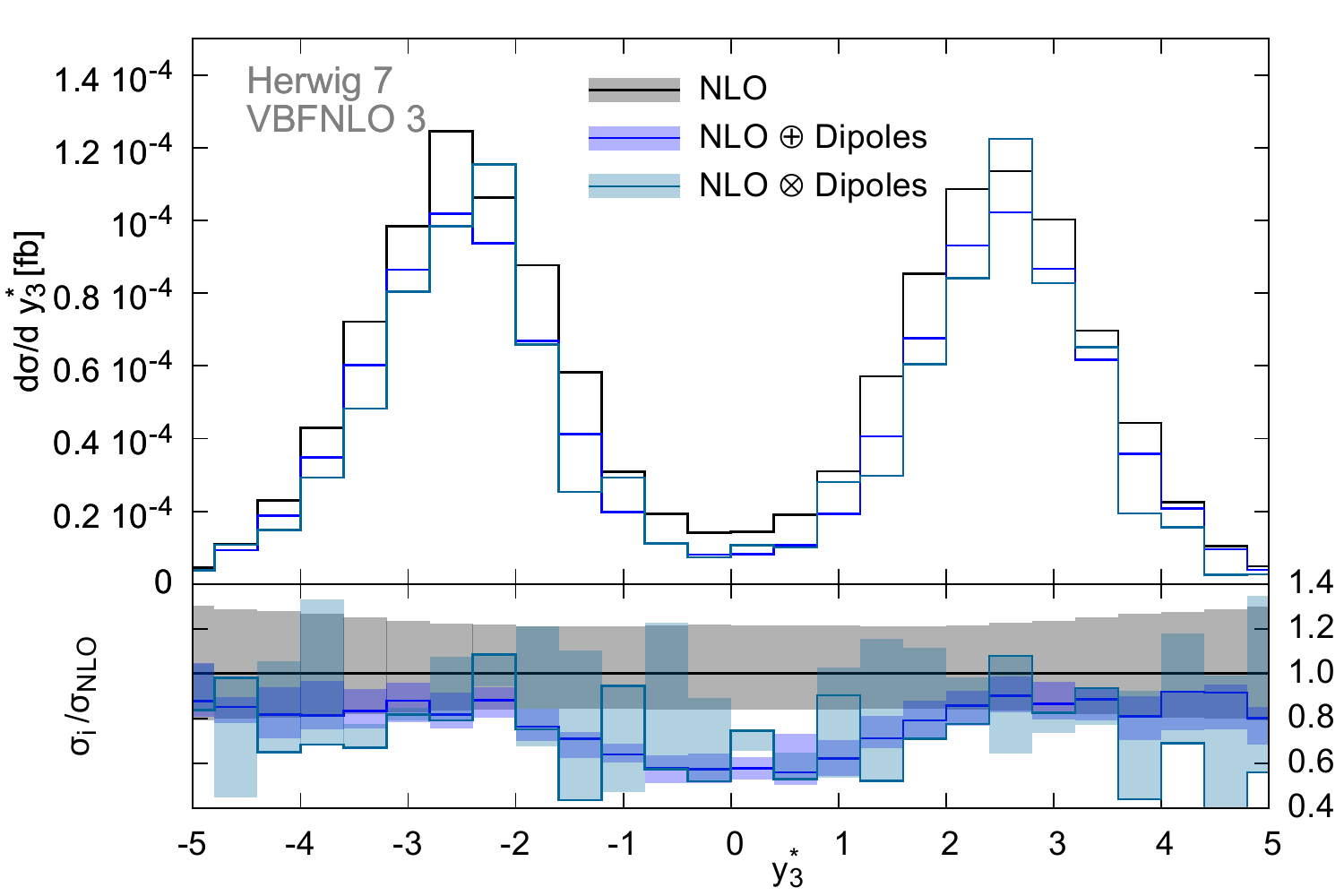} 
\hfill~
\caption{
\textit{Left:} Schematic picture of additional radiation generated by
the parton shower. \textit{Right:}
Rapidity of the third jet relative to the two tagging jets, comparing
the fixed-order NLO result and NLO combined with the dipole shower using
both subtractive and multiplicative matching.
Right panel taken from Ref.~\protect\cite{Rauch:2016upa}.
}
\label{fig:uncert_powheg}
\end{figure}
Turning to uncertainties and parton-shower matching systematics, we
present in Fig.~\ref{fig:uncert_mcatnlo} in the left panel the
invariant-mass distribution of the four leptons for calculations of
fixed-order NLO, LO plus dipole shower, as well as NLO combined with
dipole and angular-ordered shower using subtractive matching. The lower
panels show the cross sections normalised to the fixed-order result,
and variation bands for factorisation ($\mu_F$), renormalisation
($\mu_R$), and hard veto ($\mu_Q$) scale as well as the overall envelope
($\mu_{\text{tot}}$) normalised to the respective central prediction.
Individual scales are varied by a factor of $2^{\pm1}$ around the central
value, taken as the transverse momentum of the leading jet. This
dynamical scale is directly related to the momentum transfer of the
exchanged bosons, and also an important quantity in the parton-shower
algorithm, thus making it an ideal choice. For the overall band, we
allow the different scales to vary independently, provided the ratio of
any two scales is within the range $[\frac12;2]$ as well. As expected
for a quantity constructed from the electroweak subsystem, shower
effects on the shape are small, while the inclusive rate gets reduced
due to events migrating outside the allowed cut range. The transition
from LO$\oplus$PS to NLO$\oplus$PS leads to a clearly reduced scale
uncertainty. The increased renormalisation dependence of NLO$\oplus$PS
at small invariant masses is due to a larger value of the strong
coupling constant, which leads to more radiation and thus larger
migration effects.
The right panel of Fig.~\ref{fig:uncert_mcatnlo} shows the rapidity of
the third jet relative to the two tagging jets, $y^*_3 = y_3 -
\frac{y_1+y_2}2$. This quantity is purely due to shower effects in the
LO$\oplus$PS simulation, and of leading-order accuracy for all other
curves. Consequently, we observe much larger effects, both in changes
of the shape as well as bigger uncertainty bands. In particular the
LO$\oplus$PS curve exhibits a large dependency on the shower hard scale
and also a strongly reduced cross section in the central region. The
reason for this becomes clear when considering the schematic picture
shown in Fig.~\ref{fig:uncert_powheg} on the left. As the colour
correlation of the tagging jet is solely with the beam remnant, but not
with the other quark line as the exchanged boson is a colour singlet,
one expects that additional radiation is generated predominantly between
the tagging jet and the beam line, and thus for large absolute values of
rapidity. Corrections by the hard matrix element when using the NLO
calculation increase this prediction somewhat, but again both parton
showers reduce the rate in the central region compared to the
fixed-order result. This also holds for the multiplicative matching
scheme, which one can see from the comparison shown in the right panel
of Fig.~\ref{fig:uncert_powheg}. For this scheme we use the resummation
profile~\cite{Bellm:2016rhh}, and the bands in the ratio denote a joint
variation of all four scales by a factor $2^{\pm1}$. 

In summary, matching and parton-shower uncertainties are well under
control for this process.

\acknowledgments
We would like to thank the organisers for the stimulating atmosphere at
the conference. We are grateful to the other members of the Herwig and
VBFNLO collaborations for encouragement and helpful discussions.
SP acknowledges support by a FP7 Marie Curie Intra European Fellowship
under Grant Agreement PIEF-GA-2013-628739.

\end{document}